\documentclass{article}
\begin{document}
\title{ Coriolis Force and Sagnac Effect }
\author{{\bf Merab Gogberashvili}\\
 Andronikashvili Institute of Physics, 6 Tamarashvili Str.,
Tbilisi 380077, Georgia \\
{\sl E-mail: gogber@hotmail.com }}
\maketitle
\begin{abstract}
We consider the optical Sagnac effect, when the fictitious gravitational
field simulates the reflections from the mirrors. It is shown that no
contradiction exists between the conclusions of the laboratory and
rotated observers. Because of acting of gravity-like Coriolis force the
trajectories of co- and anti-rotating photons have different radii in
the rotating reference frame, while in the case of the equal radius the
effective gravitational potentials for the photons have to be different.

\medskip

\end{abstract}
%%%%%%%%%%%%%%%%%%%%%%%%%%%%%%%%%%%%%%%%%%%%%%%%%%%%%%%%
\vskip 1cm

Since its discovery at the beginning of the XX century the Sagnac effect
\cite{Sa} has play an important role in the understanding and
development of fundamental physics (for a review see \cite{St}). The
Sagnac effect is the dependence of the interference pattern of the
rotating interferometer on the direction and speed of rotation. This
phenomenon is universal and manifested for any kind of waves, including
the matter waves and has found a variety of applications for the
practical purposes and in the fundamental physics \cite{St}. It has also
a direct experimental verification for large distances in the
experiments of clocks transported around Earth \cite{AlWe}. For the
laboratory observer the Sagnac effect seems to have simple explanation -
because of the rotation the round-trip distance traveled by the waves
co-rotating with the platform is grater than for the anti-rotating
waves.

Some misunderstandings appear when considering the optical Sagnac effect
from the viewpoint of an observer on the rotating disk (see for example
\cite{Se}), where the both light beams travel the same distance. Since
the round-trip time along a closed path on the disk is different for
clockwise and anti-clockwise photons the non-inertial observer can
conclude that the speed of the light measured locally on the disk
depends on the direction and speed of rotation. On the other hand, in
accordance with General Relativity consistently defined speed of light
for any observer turns out to be exactly the same, just as in the case
of an inertial reference frame \cite{RiTa}.

In the context of the Sagnac effects the null result of the
Michelson-Morley experiment is also not clear. Applying the same logic
to Sun centered rotating frame in which Earth is fixed, one would expect
different light speeds as seen from Earth. Another problem is
explanation of the experiments where the effects of placing of
transparent media on the paths of the beams were analyzed \cite{Po}. No
interference fringe change was observed between the light beams in air
and in other media configurations. However, the change was observed when
the media was stationary and the platform was rotated.

The Sagnac effect have been attempted by several authors to explain:
using the General Relativity \cite{Po,An}; from the special relativistic
Doppler effect at the mirrors \cite{DrYa}; by coupling the momentum of
interfering particles to rotation \cite{Sak}; introducing various types
of relativistic transformations for rotating frames \cite{TTHB}; using
the restricted formula of the 3-space element \cite{Ni}; considering the
non-invariance of the light speed for the rotating observers
\cite{Se,RiTaPeKlPe}.

To make a round-trip on the disk the photon must be reflected by several
mirrors and its path has the shape of a polygon. For example, in the
classical Sagnac experiment \cite{Sa} the light path was quadratic. The
form of the trajectory is close to the circle when the number of sides
of the polygon is large. In this paper we would like to introduce the
fictitious gravitational potential and write the effective metric for
the photons in the form
\begin{equation} \label{ds1}
ds^2 = c^2(1-2A/r)dt^2 - dr^2 - r^2d\phi^2 -dz^2 .
\end{equation}
The "gravitational mass" $A$ is a constant simulating the reflection of
photons. In order to have a closed trajectory of photons in (\ref{ds1})
we do not change the spatial part of the metric.

The geodesic equation
$dk^\alpha /d\lambda + \Gamma_{\mu\nu}^{\alpha}k^\mu k^\nu = 0$
for the photons moving on the circle ($z = 0, r = \rho = const $)
reduces to
\begin{equation} \label{system1}
\frac{dt}{d\lambda} = const , ~~~~~ \frac{d\phi}{d\lambda} = const ,
~~~~~
\frac{c^2 A}{r^2}\left(\frac{dt}{d\lambda}\right)^2 = r
\left(\frac{d\phi}{d\lambda}\right)^2 ,
\end{equation}
where $k^\nu = dx^\nu/d\lambda $ is a wave 4-vector and $\lambda $ is
some parameter varying along the ray.

This system and the Hamilton-Jacobi equation for the photons
$ g_{\mu\nu}k^\mu k^\nu = 0 $,
which is equivalent to $ds^2 = 0$, gives the value of the radius of the
trajectory $\rho = 3A$, and the expression for the angular velocity
$d\phi /dt = (c^2A/\rho^3)^{1/2} $ as well. Finally we find the expected
value of the proper time $\tau = \sqrt{g_{tt}}t = t/\sqrt{3} = 2\pi \rho
/ c $ which is needed by the photons to travel the circle of the radius
$\rho $.

For the case of disks uniformly rotating in different directions with
the angular velocities $\pm \omega$ the metric (\ref{ds1}) transforms to
\cite{LaLi}
\begin{equation} \label{ds2}
ds_\pm^2=c^2(1-2A/r - r^2\omega ^2/c^2) dt^2-dr^2-r^2d\phi ^2 \pm
2r^2\omega d\phi dt -dz^2 .
\end{equation}
Transformation of (\ref{ds1}) to (\ref{ds2}) is in the full accordance
with numerous cyclotron experiments supporting the contention that
(\ref{ds2}) is the correct form of the metric of a rotating disk, but
only for the inertial observer (despite of the fact that it corresponds
to the Galilean composition law of velocities \cite{KiKr}).

The geodesic equations (\ref{system1}) now have the form
\begin{equation} \label{system2}
\frac{dt}{d\lambda} = const , ~~~~~ \frac{d\phi}{d\lambda} = const ,
~~~~~
\left( \frac{c^2 A}{r^2} - r^2\omega^2
\right)\left(\frac{dt}{d\lambda}\right)^2 = r
\left(\frac{d\phi}{d\lambda}\right)^2 \mp
2r\omega\frac{dt}{d\lambda}\frac{d\phi}{d\lambda} .
\end{equation}
This system and the Hamilton-Jacobi equation, yields $\rho = 3A$ and
$d\phi/dt = (c^2A/\rho^3)^{1/2} \pm \omega $. In the first approximation
in $(\omega\rho /c)$ the last relation gives the travel times $t \approx
\tau \mp 2\pi \rho^2\omega /c^2$ of co- and anti-rotated photons along
the circle of the radius $\rho $. For the time lag we recover the
well-known Sagnac formula
\begin{equation} \label{delta1}
\delta \tau \approx 4 S \omega /c^2 ,
\end{equation}
where $S = \pi \rho^2$ is area of the projection of the trajectory
spanned by the photons.

As in every stationary field, the clocks on the rotating body cannot be
uniquely synchronized at all points and the metric (\ref{ds2}) can not
be used for a rotating observer. In the rotating frame two fictitious
gravity-like forces appear, namely the centrifugal and Coriolis forces.
This is an illustration of the equivalence principle, which asserts that
gravity and the accelerated motion are locally indistinguishable. The
centrifugal force does not affect photons. However, it causes the
deformations of the mirrors and shifts the relaxation times for the
electrons in the conductivity zone, which are forming the mirrors. Thus
the centrifugal force changes the constant $A$. In the coordinates of
the observer on the disk (\ref{ds1}) can be written in the form
\begin{equation} \label{ds3}
ds^2 = c^2(1-2B/R)dT^2 - dR^2 - R^2d\Phi^2 - dZ^2 .
\end{equation}
The "mass" $B$ simulates the motion of the photons along the circle of
the radius $P$, just as in the inertial case.

The motion of the photons under the influence of the Coriolis force
\cite{Sak} can be described by the following equation
\begin{equation} \label{coriolis1}
d\vec{k}/dT = 2 [\vec{k} \vec{\Omega }] ,
\end{equation}
where $\vec{\Omega} $ is some 3-vector which appears in a rotated
reference system. The Coriolis force acting on the tangential photons
co-rotating (or anti-rotating) with the disk is directed radially
outward away from (or inward towards) the centre. It means that the
photons are deflected out (or to) the centre as if they are
gravitationally repelled (or attracted) by the disk. The result is that
the light beam reflected from one mirror on the disk to the next one
falls on the point that located before (or next) the one it was falling
in the non-rotating case. Thus for a time interval measured by the clock
at the mirror the photons cover less (or more) distances along the disk
circumference, in the full agreement with the inertial observer.

For the circular motion the Coriolis force has only the radial component
and (\ref{coriolis1}) yields
\begin{equation} \label{coriolis2}
dk^\Phi/dT = \pm 2 R k^\Phi \Omega .
\end{equation}
Then the geodesic equations of photons moving along the circular orbits
($Z = 0, R = P = const $)
have to be written in the form different from (\ref{system1})
\begin{equation} \label{system3}
\frac{dT}{d\Lambda} = const , ~~~~~ \frac{d\Phi}{d\Lambda} = const ,
~~~~~
\frac{c^2 B}{R^2}\left(\frac{dT}{d\Lambda}\right)^2 = R
\left(\frac{d\Phi}{d\Lambda}\right)^2 \pm 2R\Omega
\frac{d\Phi}{d\Lambda}\frac{dT}{d\Lambda} .
\end{equation}
This system and the Hamilton-Jacobi equation leads to
\begin{equation} \label{P1}
3B = P \pm 2\Omega P^2/c , ~~~~~ d\Phi/dT = c/P \approx c/3B \pm \Omega
.
\end{equation}
>From the first equation we see that for the observer on the disk the
radii of the trajectories for clockwise and anti-clockwise photons are
different
\begin{equation} \label{P2}
P_\pm = \pm\frac{c}{4|\Omega|}\left(\sqrt{1 \pm \frac{24B|\Omega|}{c}} -
1\right) \approx 3B \left( 1 \mp \frac{3B\Omega}{c} \right) .
\end{equation}
>From (\ref{P1}) it follows that time intervals
$T_\pm = 2\pi 3B/c \mp 2\pi(3B)^2\Omega/c^2 $
needed by the beams to travel around the disk are different and give the
time lag (\ref{delta1}), where now $S = \pi P^2$. In order to have the
same radius for the beams (this corresponds to $P_+ = P_-$ in
(\ref{P2})) in the metric (\ref{ds3}) different "gravitational masses"
$B_\pm$ have to be introduced. From (\ref{P2}) one obtains $B_+ - B_-
\approx 6B^2\Omega /c $ and the second equation in (\ref{P1}) gives
(\ref{delta1}) again.

In the framework of the tower experiment \cite{PoSn} the action of the
Coriolis force on the light is equivalent to reception of co-rotated
photons by a mirror from a higher gravitational potential (from the
upstairs) and of anti-rotated photons from the lower gravitational
potential (from the downstairs), since after the round reflection the
cumulate angle factor disappears. For the mirror this effect looks like
a redshift of the anti-rotated and a blueshift of the co-rotated photons
as compared with the inertial ones.

The similar situation takes place for the case of the gravitational
clock experiment \cite{AlWe} where two identical clocks were flown
around Earth in different directions. The gravitational blue shift and
the special relativistic slowing of the moving clocks are background
affects and are the same for both of flying clocks. Only the Coriolis
force distinguishes between the co- and anti-rotating senses of motion.
For the co-rotated case the 'repulsive' Coriolis force tend to offset
the gravitational attraction effect and the clock is in a lower
gravitational potential as seen by the Earth observer. For the
anti-rotating case the gravitational and Coriolis forces have the same
directions. It is well known that the clocks in a higher gravitational
potential tick faster. As the result the anti-rotating clock will be
ahead relative to the co-rotating one with the time shift
(\ref{delta1}). The detailed description of gravitational clock effect
for the inertial observer is done in \cite{Fe}.

In the Michelson-Morley experiment the light makes a round trip along
Earth surface, but along one half of the way it co-rotates and then
anti-rotates with Earth. Therefore, the effects of 'repulsive' and
'attractive' Coriolis forces offset each other. One can also understand
the effect of the transparent media on the light beams. The co-rotating
medium just changes the "gravitational mass" $A$ in (\ref{ds2}) and does
not affect (\ref{delta1}), while the inertial medium changes the
Coriolis force and the time lags as well.

\medskip
{\bf Acknowledgements:} Author would like to acknowledge the hospitality
extended during his visits at the Theoretical Divisions of CERN and of
Abdus Salam International Centre for Theoretical Physics where this work
was done.


\begin{thebibliography}{99}

\bibitem{Sa} G. Sagnac,
             {\it Compt. Rend. Acad. Sci. Paris}, {\bf 152}, 310 (1911);

              ibid, {\bf 157}, 708 (1913).

\bibitem{St} G. E. Stedman,
             {\it Rep. Prog. Phys.}, {\bf 60}, 615 (1997).

\bibitem{AlWe} D. W. Allan, M. A. Weiss,
               {\it Science}, {\bf 228}, 69 (1985).

\bibitem{Se} F. Selleri,
             {\it Found. Phys. Lett.}, {\bf 10}, 73 (1997).

\bibitem{RiTa} G. Rizzi, A. Tartaglia,
               {\it Found. Phys.}, {\bf 28}, 1663 (1998).

\bibitem{Po} E. J. Post,
             {\it Rev. Mod. Phys.}, {\bf 39}, 475 (1967).

\bibitem{An} J. Anandan,
                  {\it Phys. Rev}, {\bf D 24}, 338 (1981).

\bibitem{DrYa} M. Dresden and C. N. Yang,
               {\it Phys. Rev.}, {\bf D 20}, 1846 (1979).

\bibitem{Sak} J. J. Sakurai,
              {\it Phys. Rev.}, {\bf D 21}, 2993 (1980).

\bibitem{TTHB}  M. Trocheris,
                {\it Phil. Mag.}, {\bf 40}, 1143 (1949); \\
                H. Takeno,
                {\it Prog. Theor. Phys.}, {\bf 7}, 367 (1952); \\
                L. Herrera,
                {\it Nuovo Cim.}, {\bf B 115}, 307 (2000); \\
                V. Bashkov and M. Malakhaltsev,
                gr-qc/0011061.

\bibitem{Ni}  H. Nikolic,
               {\it Phys. Rev.}, {\bf A 61}, 032109 (2000).

\bibitem{RiTaPeKlPe} A. Tartaglia,
                {\it Phys. Rev.}, {\bf D 58}, 064009 (1998); \\
                A. Peres,
                {\it Phys. Rev.}, {\bf D 18}, 2173 (1978); \\
                R. D. Klauber,
                gr-qc/0103076; \\
                V. Petkov,
                gr-qc/9909081.

\bibitem{LaLi} L. Landau and E. Lifshitz,
               {\it The Classical Theory of Fields} (Pergamon Press,
1962).

\bibitem{KiKr} S. Kichenassamy and P. Krikorian,
                  {\it J. Math. Phys.}, {\bf 35}, 5726 (1994).

\bibitem{PoSn} R. V. Pound and J. L. Snider,
                  {\it Phys. Rev.}, {\bf B 140}, 788 (1965).

\bibitem{Fe} R. Feynman,
               {\it Lectures on Gravitation} (Addison-Wesley Publ.,
1995).

\end{thebibliography}
\end{document}